\documentclass[namedreferences]{SolarPhysics}
\usepackage{epsfig}
\usepackage{graphicx}

\begin{document}
\begin{article}
\begin{opening}

\title{GEOMETRICAL AND PHYSICAL PROPERTIES OF SXR LOOP-TOP FLARE KERNELS}

\author{P. \surname{PRE\'S} and S. \surname{KO{\L}OMA\'NSKI}}

\runningauthor{PRE\'S AND KO{\L}OMA\'NSKI}
\runningtitle{Geometrical and Physical Properties of SXR Flare Kernels}

\institute{Instytut Astronomiczny Uniwersytetu Wroc{\l}awskiego,
             Kopernika 11, 51-622 Wroc{\l}aw, Poland\\
             \email{pres,kolomans@astro.uni.wroc.pl}}

\date{Received ; accepted }

\begin{abstract}

We investigate how the geometrical and physical properties of soft X-ray
flare kernels change with their altitude above the photosphere.
We analyze limb flares well observed by {\em Yohkoh}/SXT showing clear
geometry with well separated loop-top kernels. Our analysis concerns relations
between kernel size, plasma pressure, energy release and the kernel
altitude.
We define scaling laws describing how the sizes and its physical
properties of kernels vary with the altitude above the photosphere. We interpret
the observed relations in terms of the general magnetic structure of
active regions.

\end{abstract}

\end{opening}

\section{Introduction}

The loop structures making up soft X-ray (SXR) flares generally show
bright regions at their tops, which we here call kernels. They form
before maximum and in long-duration flares may last for hours, with new
kernels forming after the flare maximum. A kernel's surface brightness
may be ten times as bright as the rest of the flare loop.
High-resolution images show the thermal structure to be uniform (Feldman
et~al., 1995; Jakimiec et al., 1998) and about half the
total flare emission measure is contained with the kernel.

The classic model describes solar flare as a hydrodynamic reaction of
dense chromospheric plasma to a sudden release of energy in the corona.
This energy conducted down along the magnetic lines and transfered by
energetic electrons heats up the cool, dense chromosphere and forces it
to fill magnetic flux-tube, where the energy release occurred (Bentley
et al., 1994; Tomczak, 1997). In a case of simple magnetic structure the
plasma should fill it uniformly, without local condensation. The
observed kernels can be brighter due to higher density or temperature,
and in a simple magnetic loop any kernel should disappear in a very
short time due to expansion or conduction. Some kind of restriction
efficiently preventing outflow of mass and energy even for hours must be
present at the boundary of a kernel \cite{vorpahl}. 

Loop-top kernels are recognizable in the the flare images from {\em
Skylab}. Although widely commented on since the {\em Yohkoh} Soft X-ray
Telescope (SXT) observations (e.g. Acton et~al., 1992; Doschek, Strong
and Tsuneta, 1995; Jakimiec et~al., 1998), they are still not well
understood. Recently proposed mechanisms are based on MHD turbulence
\cite{jakimiec02j}, fast-mode MHD shock \cite{yokoyama} or magnetic trap
at the top of the cusp structure \cite{karlicky}. 

The ratio of the kernel soft X-ray emission to the total emission of the
flare is in the range 0.35 to 0.45 for different flares, and moreover
remains nearly constant for the whole period of kernel existence
\cite{doschek}. This property makes the kernels a
usefull tool to analyze the evolution of the whole flare. In this paper
we focus on how the geometrical and physical properties of loop-top
kernels depend on their altitude above the photosphere. To this work we
utilize the observations of {\em Yohkoh} SXT instrument \cite{sxt},
which form a large database of soft X-ray flares over the
period from October 1991 to December 2001 (maximum of Cycle 22 to
maximum Cycle 23). Data from the SXT allow us to determine at the same
time both geometrical properties of flare kernels as well as physical
properties of the plasma.

\section{Sample description}

To estimate a kernel altitude ($h$) we need to know its position on the
SXT image and assume above which point of the photosphere it is situated
(sub-kernel point). The H$\alpha$ database and `Masuda method' \cite{masuda}
are the most helpful in locating this point. The
uncertainty of its position is the main source of error in estimating
the kernel altitude. The uncertainty is least at the limb, so we chosed
to analyze events close to the limb. We searched the GOES database for
flares with location within 10 degrees from the limb. Looking at SXT
images of these flares we chose events with simple structure showing a
single loop-top kernel or in the case of multi-kernel flares we chose
the kernels well separated from other sources. We enlarged the sample
with some flares at longitudes slightly grater than 10 degrees from the
limb or located behind the limb. To obtain reliable physical properties
of the emitting plasma we selected the flares observed with the use of
the SXT thick aluminium (Al12) and beryllium filters (Be119) in full
angular resolution ($2.\! ''45$) during the flare maximum.


\begin{table*}
\caption[]{The sample of selected flares. Remarks in the last column: 
A - flare with one kernel, B - flare with more 
than one kernel but only one analyzed, C - flare with two 
kernels, both analyzed.}
\label{t1}
\centering
\begin{tabular}{c | c | c   c | c   c | c}
\hline
No & date & \multicolumn{2}{c|}{GOES} & \multicolumn{2}{c|}{location}     & remarks \\ 
   &      & maximum [UT]& X-ray class & co-ordinates & NOAA active region &         \\ \hline 
1  & 17-Nov-91 & 07:16 & M1.1 & E87 ~ ~ N13 & 6929 & A \\
2  & 19-Nov-91 & 09:32 & C8.5 & W64 ~ ~ S13 & 6919 & A \\
3  & 02-Dec-91 & 05:01 & M3.6 & E87 ~ ~ N16 & 6952 & B \\
4  & 09-Dec-91 & 09:44 & M4.1 & E89 ~ ~ S05 & 6966 & A \\
5  & 09-Dec-91 & 23:49 & M1.0 & E81 ~ ~ S07 & 6966 & A \\
6  & 13-Jan-92 & 17:34 & M2.0 & W90 ~ ~ S16 & 6994 & A \\
7  & 30-Jan-92 & 17:15 & M1.6 & E83 ~ ~ S13 & 7042 & A \\
8  & 06-Feb-92 & 03:29 & M7.6 & W85 ~ ~ N06 & 7030 & C \\
9  & 17-Feb-92 & 15:46 & M1.9 & W80 ~ ~ N15 & 7050 & A \\
10 & 28-Jun-92 & 05:14 & X1.8 & W102 ~ ~ N13& 7205 & B \\
11 & 28-Jun-92 & 14:24 & M1.6 & E90 ~ ~ N14 & 7216 & B \\
12 & 05-Jul-92 & 12:04 & C4.1 & E82 ~ ~ S11 & 7220 & A \\
13 & 12-Oct-92 & 21:53 & C2.5 & W83 ~ ~ S19 & 7303 & A \\ 
14 & 02-Nov-92 & 03:08 & X9.0 & W99 ~ ~ S24 & 7321 & A \\
15 & 21-Nov-92 & 07:13 & C5.0 & W81 ~ ~ S16 & 7341 & A \\
16 & 23-Nov-92 & 13:59 & C4.0 & W84 ~ ~ S08 & 7342 & A \\
17 & 29-Nov-92 & 08:58 & C9.1 & W90 ~ ~ S27 & 7345 & A \\
18 & 17-Feb-93 & 10:40 & M5.8 & W88 ~ ~ S07 & 7420 & A \\ 
19 & 02-Mar-93 & 15:10 & C5.0 & E82 ~ ~ S04 & 7440 & A \\
20 & 15-Mar-93 & 21:35 & M2.9 & W94 ~ ~ S02 & 7440 & A \\
21 & 12-Jun-93 & 09:06 & C3.5 & W95 ~ ~ S11 & 7518 & A \\
22 & 25-Jun-93 & 03:22 & M5.1 & E84 ~ ~ S09 & 7530 & A \\
23 & 27-Sep-93 & 12:12 & M1.8 & E86 ~ ~ N10 & 7590 & A \\
24 & 28-Jan-94 & 17:04 & M1.8 & W87 ~ ~ N06 & 7654 & A \\
25 & 27-Feb-94 & 09:20 & M2.8 & W97 ~ ~ N10 & 7671 & C \\
26 & 30-Aug-94 & 19:54 & C6.2 & E77 ~ ~ S09 & 7773 & A \\
27 & 21-Apr-95 & 13:41 & C5.1 & W75 ~ ~ S01 & 7863 & A \\
28 & 17-Sep-97 & 11:43 & M1.7 & W81 ~ ~ N21 & 8084 & A \\
29 & 26-Nov-97 & 04:47 & C4.7 & E86 ~ ~ N20 & 8113 & A \\
30 & 08-May-98 & 02:04 & M3.1 & W88 ~ ~ S16 & 8210 & A \\
31 & 18-Aug-98 & 08:24 & X2.8 & E91 ~ ~ N33 & 8307 & A \\
32 & 18-Aug-98 & 22:19 & X4.9 & E86 ~ ~ N33 & 8307 & A \\
33 & 19-Aug-98 & 14:26 & M3.0 & E79 ~ ~ N33 & 8307 & A \\
34 & 22-Nov-98 & 06:42 & X3.7 & W76 ~ ~ S28 & 8384 & A \\
35 & 22-Nov-98 & 16:23 & X2.5 & W81 ~ ~ S28 & 8384 & A \\
36 & 24-Nov-98 & 02:20 & X1.0 & W98 ~ ~ S29 & 8384 & A \\
37 & 23-Dec-98 & 06:59 & M2.3 & E90 ~ ~ N23 & 8421 & B \\
38 & 25-Jul-99 & 13:38 & M2.4 & W82 ~ ~ N39 & 8639 & B \\
39 & 21-Sep-99 & 10:47 & C6.4 & W96 ~ ~ S25 & 8692 & A \\
40 & 23-May-00 & 17:54 & C4.3 & W75 ~ ~ S21 & 8996 & B \\
41 & 01-Jun-00 & 06:17 & M2.5 & E80 ~ ~ N21 & 9026 & A \\
42 & 30-Sep-00 & 23:21 & X1.2 & W93 ~ ~ N08 & 9169 & A \\
43 & 14-Oct-00 & 08:40 & M1.1 & W81 ~ ~ N02 & 9182 & A \\
44 & 14-Oct-00 & 12:05 & C8.4 & W80 ~ ~ N02 & 9182 & A \\
45 & 28-Oct-00 & 07:10 & C9.7 & E80 ~ ~ N08 & 9212 & A \\
46 & 14-Nov-00 & 16:34 & M1.0 & E82 ~ ~ N13 & 9233 & A \\
47 & 11-Mar-01 & 08:56 & C5.0 & E82 ~ ~ S15 & 9376 & A \\
48 & 29-Oct-01 & 01:59 & M1.3 & W88 ~ ~ N13 & $--$ & A \\
\hline
\end{tabular}
\end{table*}

As each flare evolves greatly with time, we compared the properties of
flare kernels at the time of maximum emission in the SXT Al12 filter.
We found 48 suitable events (see Table~\ref{t1}) during the entire
period of {\em Yohkoh} operations (Oct 1991 to Dec 2001). The sample consists 
of 8 X-class flares, 24 M-class and 16 C-class events. 34 of them were
located within 10 degrees from the limb, while for 6 flares the SXR
footpoints seen on SXT images were located between 60 and 80 degrees of
heliographical longitude. To this number we added 8 flares located
no more than 12 degrees behind the limb for which the loop-top kernels
were not occulted by the limb.  In our sample only two flares had two
well-separated loop-top kernels, allowing us to analyze both kernels. 6
flares had more than one kernel but only one of them could be analyzed.
The rest of our events (i.e. 40) had only one kernel. Thus, a total of
50 flare kernels were analyzed. 

\section{Analysis}

The geometrical parameters of interest here are the size and altitude of
each kernel, and the physical ones are plasma density, pressure and
energy release rate in the kernel. The kernel size we defined by a 50\%
isophote ($I_{50}$) relative to the brightest pixel in the Al12 image.
We measured the area, $A$, of the kernel projected on the plane of the
sky. The kernel altitude, $h$, was calculated from the position of its
centroid and coordinates of the sub-kernel point. The kernel centroid
was estimated as a center of gravity in the sense of intensity
distribution within the isophote $I_{50}$. From the kernel area we
summed the Al12 and Be119 flux to estimate the emission measure, {\em
EM}, and temperature, $T$, of emitting plasma using the filter ratio
method as described by Hara et~al. (1992).

The estimation of kernel volume is not straightforward. Usually the kernel image
has a circular or elliptical shape. If we assume that its volume is
ellipsoidal we can calculate it as $V=(4/3)\,\pi abc$ where $a$, $b$ and
$c$ are the three semi-axes. In general we can estimate two semi-axes
from the kernel image, but the third dimension along the line-of-sight must be
reasonably guessed. Generally the smaller of the two taken from
the image is used to estimate the volume. This assumes that the main
axis of ellipsoid is located in the plane of image. However, we do not
know definitely the orientation of the kernel relative to the plane of the sky,
so the effective depth of the kernel can be greater. To avoid this
underestimation we assume that the depth is close
to the geometrical mean of the two axes seen on the image. This
assumption gives the estimation of kernel volume as
$$V=\frac{4}{3}\frac{A^{3/2}}{\sqrt\pi},$$ where $A$ is the plane-of-sky
kernel area.

Having estimates of the kernel volume we can calculate the plasma mean
density and pressure within the kernel as $N_e = \sqrt{EM/V}$ and $p_e =
2 k_BTN_e$ where $k_B$ is Boltzmann constant. One should remember that
such estimations are sensitive to the influence of filling factor and in
general should be treated as lower estimates.

The kernel area is defined by half the intensity of its brightest
pixel, so the uncertainty in this pixel's flux directly affects the
uncertainty $\Delta I_{50}$ of the isophote $I_{50}$ and the uncertainty 
$\Delta A$ of the kernel area. Estimation of $\Delta A$ is also affected by the data
pixelation. For small kernels the pixelation makes it impossible to
estimate $\Delta A$, because there may be no change of $A$ within the
range ($I_{50}-\Delta I_{50}$, $I_{50}+\Delta I_{50}$). To overcome this
limitation we closely analyzed for each kernel how the number of pixels
brighter than a given level changes with an isophote $I$ in the range
($I_{50}-5\,\Delta I_{50}$, $I_{50}+5\,\Delta I_{50}$). The step function 
obtained we fitted by a low degree polynomial. Using this
polynomial we interpolated the area $A$ at the $I_{50}$ level and
estimated the error $\Delta A$ for $\Delta I_{50}$. In the same way we
calculated the flux emitted from within the isophote $I_{50}$ in both
SXT filters and an error of the flux. Our method allowed us to estimate
the size of a kernel and its flux with precision greater than one pixel.

The error $\Delta A$ affects not only the estimated volume but also all
physical properties of kernel plasma: the total brightness in both
filters, and therefore the kernel temperature and emission measure,
electron density and pressure. The error of kernel altitude comes from
two uncertainties, one related to the kernel centroid position and
second to the location of sub-kernel point. The second uncertainty is
usually several times larger than the first. All errors mentioned affect
the calculation of heating rate within the kernel.

\section{Results}

\begin{figure}
\resizebox{\hsize}{!}
{\includegraphics{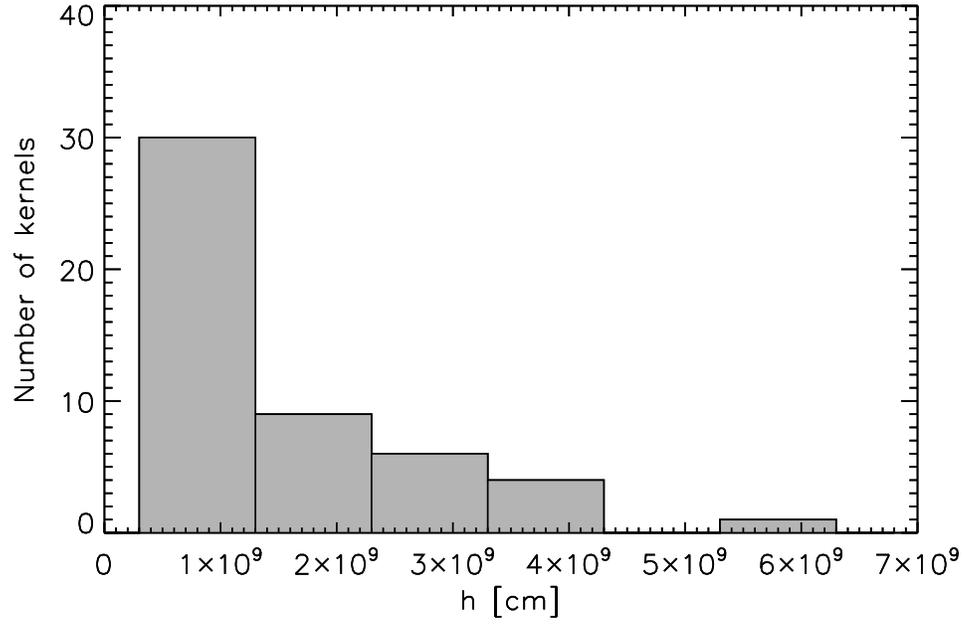}}
\caption{The observed distribution of loop-top kernel altitudes.}
\label{altitudes}
\end{figure}

The flare kernel altitudes in our sample vary from $3.2\times 10^8$ to
$5.8\times 10^9$ cm. The distribution of altitudes decreases with rising
altitude. This distribution is shown in Fig. \ref{altitudes} and
may be described as exponential, $N \propto \exp(-h/h_0)$, where $h_0 =
(2.2\pm 0.6)\times 10^9$ cm.

\begin{figure}
\resizebox{\hsize}{!}
{\includegraphics{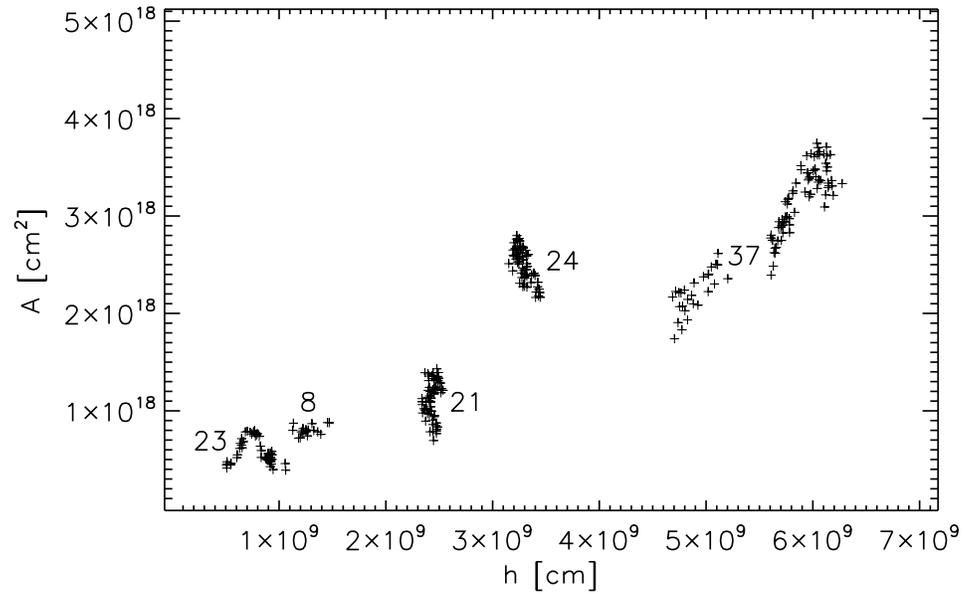}}
\caption{The changes of the projected kernel area and its altitude above 
the photosphere for a set of arbitrary selected flares. Individual kernels  
show different correlations but the overall tendency is visible. The numbers 
denote the flares listed in Table~\ref{t1}.}
\label{individual.kernels}
\end{figure}

\begin{figure}
\resizebox{\hsize}{!}
{\includegraphics{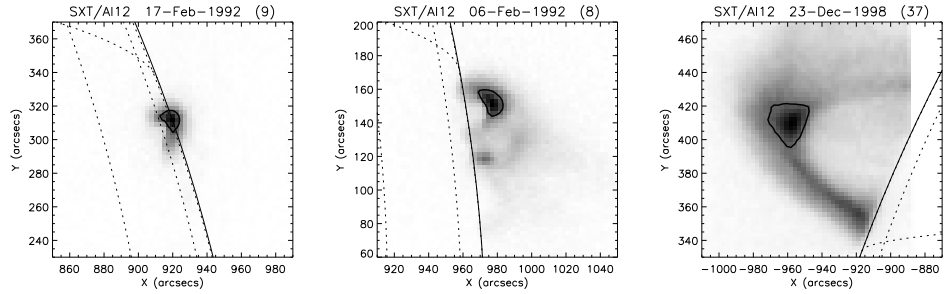}}
\caption{Three examples of flares with loop-top kernels at different
altitudes. The extension of each kernel is marked by a 50\% isophote relative to the brightest pixel in SXT Al.12 image.
Each image covers the area of $140''\times 140''$.
Note the increase of the kernel size when going to higher events.}
\label{three.flares}
\end{figure}

\begin{figure}
\resizebox{\hsize}{!}
{\includegraphics{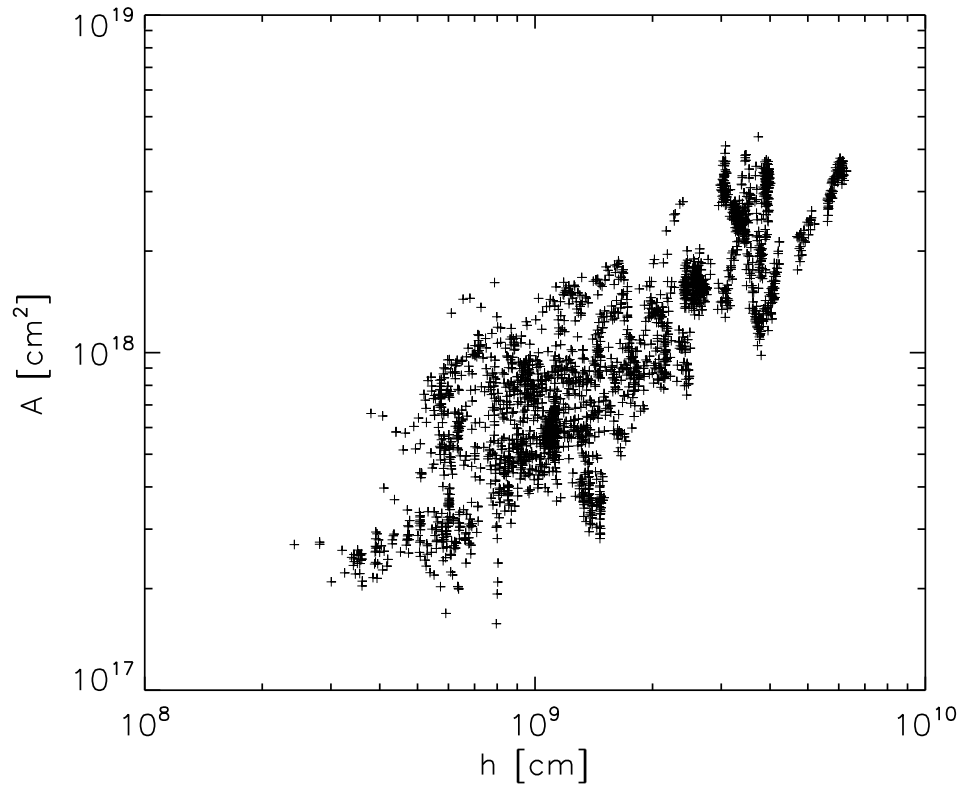}}
\caption{The changes of the projected kernel area and its altitude above 
the photosphere. Measured kernel areas and altitudes at all times during all the flares analyzed are plotted here.
The overall tendency is clearly visible, higher kernels are usually bigger.}
\label{all.kernels}
\end{figure}

\begin{figure}
\resizebox{\hsize}{!}
{\includegraphics{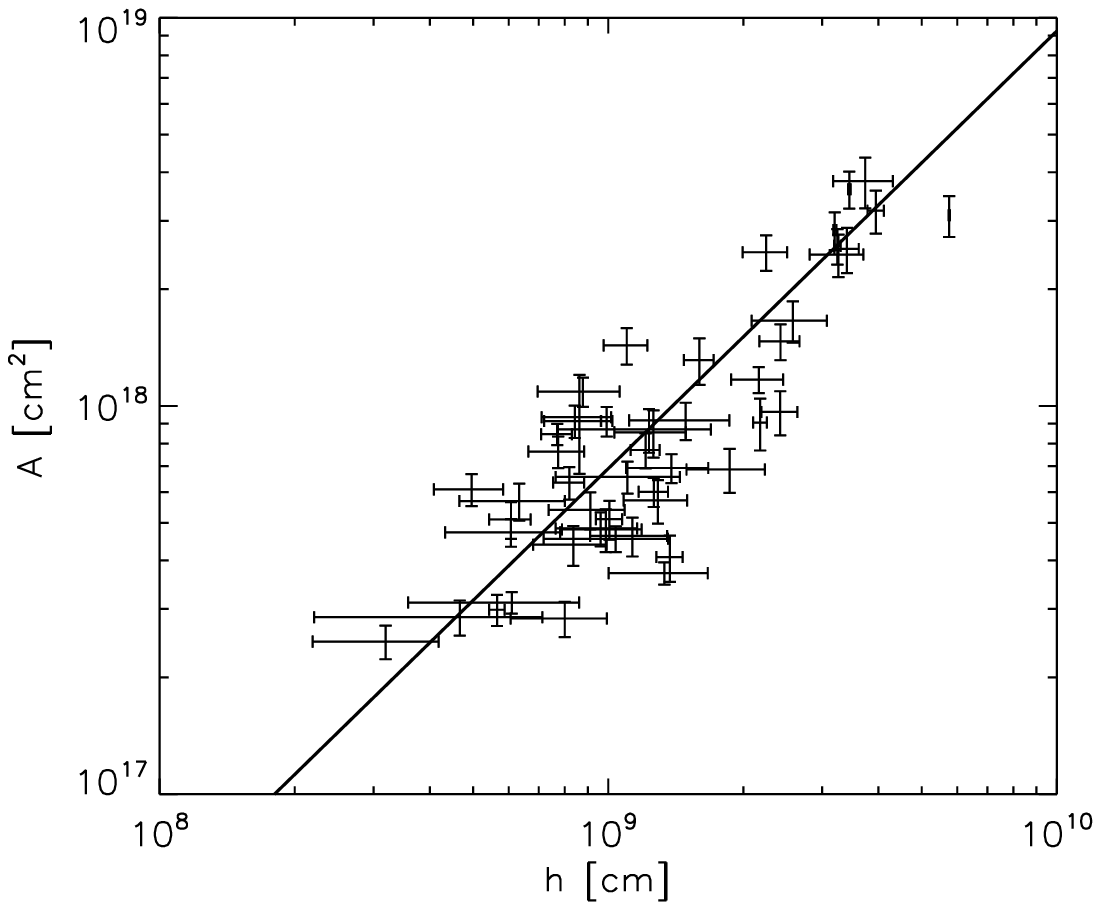}}
\caption{Relation between the projected kernel area and its altitude above
the photosphere for the time of maximum kernel brightness in the SXT Al12 filter. 
The line fitted is the power-law with slope $1.13\pm 0.04$.}
\label{a.vs.h}
\end{figure}

Figure \ref{individual.kernels} presents the relation of the kernel
size and altitude during the course of a few arbitrarily selected flares. 
As can be seen, individual flares may evolve in this diagram in very different
ways, showing both correlation and anti-correlation as well as lack of any
relationship. However, when we compare all the flares in our sample the
general tendency becomes clear: the higher the kernel, the larger its 
size. Figure \ref{three.flares} illustrates this effect in an example of three arbitrary selected flares.
The changes of kernel size and
altitude during the course of all analyzed flares are shown in Fig. \ref{all.kernels}. To
correctly compare all kernels we had to choose one specific moment in
their evolution, which we chose the moment of kernel brightness
maximum. Figure \ref{a.vs.h} presents the result of our analysis.
The projected area of a loop-top kernel distinctly rises with its
altitude. The relation is a power-law, $A\propto h^n$, where the index
$n=1.13\pm 0.04$. This relation has power index 0.56 for the
kernel mean radius, and 1.69 for the volume.

\begin{figure}
\resizebox{\hsize}{!}
{\includegraphics{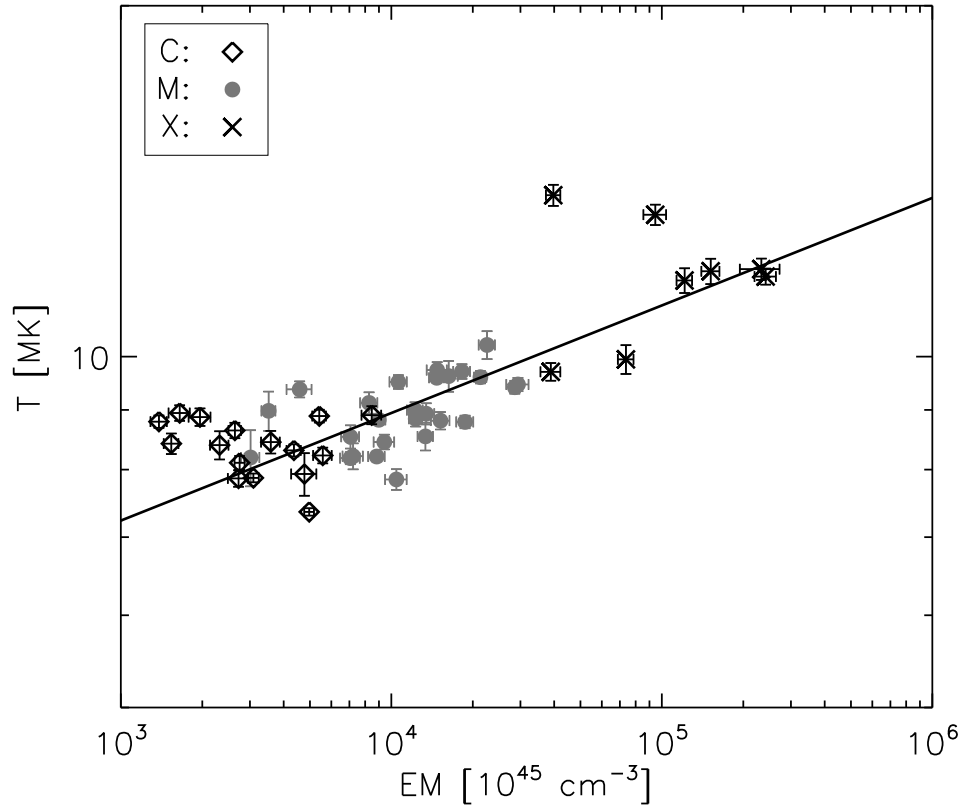}}
\caption{Observed relation between the mean kernel temperature and emission measure. 
The fitted line has slope $0.09\pm0.02$. Here and in following figures we apply different symbols for
flares of different GOES class.}
\label{t.vs.em}
\end{figure}

Our sample shows no relation between temperature or emission measure and
the altitude of the kernel. Nor is there correlation with kernel area.
Kernel emission measure correlates well with GOES X-ray class. Its
median value rises from $3.1\times10^{48}\ {\rm cm^{-3}}$ for C-class flares through
$1.2\times10^{49}\ {\rm cm^{-3}}$ for M-class flares to $1.2\times10^{50}\ {\rm
cm^{-3}}$ in X-class events. The dependence of kernel temperature with
GOES class is weaker but also evident in our sample. This relation has
been often reported and discussed in the literature in similar form as the
$T-EM$ relation. In our sample it can be well described by power-law
relation $T\propto EM^{0.09\pm 0.02}$ (see Fig.~\ref{t.vs.em}). This
dependence is weaker than others mentioned in literature. The reported
power-law slopes vary from 0.13 \cite{shibata} to
0.23 (equivalent of the exponential relation in Feldman et~al.
(1996)). This weak dependence in our sample is because of the small
sensitivity of SXT to hot plasma \cite{jakim-ker}.
Reconstructions of differential emission measure for flares generally show
the presence of two components. One component contains the relatively
cool plasma with temperatures between 5 MK and 10 MK and the other
component contains the hotter plasma with T between 15 MK and 25 MK
\cite{ania}. In the presence of the cooler component,
the hotter plasma only weakly contributes to the SXT signal.



\begin{figure}
\resizebox{\hsize}{!}
{\includegraphics{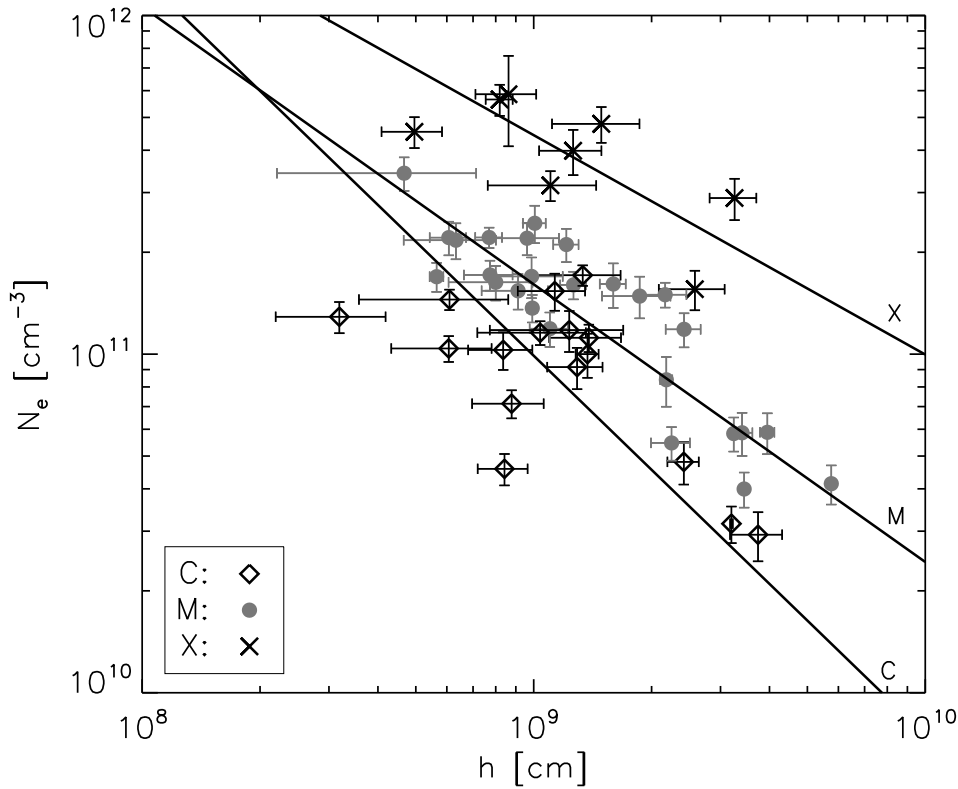}}
\caption{Relation between the kernel plasma pressure and altitude
above the photosphere. The lines show power-law fits to each of three GOES sub-samples.}
\label{n.vs.h}
\end{figure}

\begin{figure}
\resizebox{\hsize}{!}
{\includegraphics{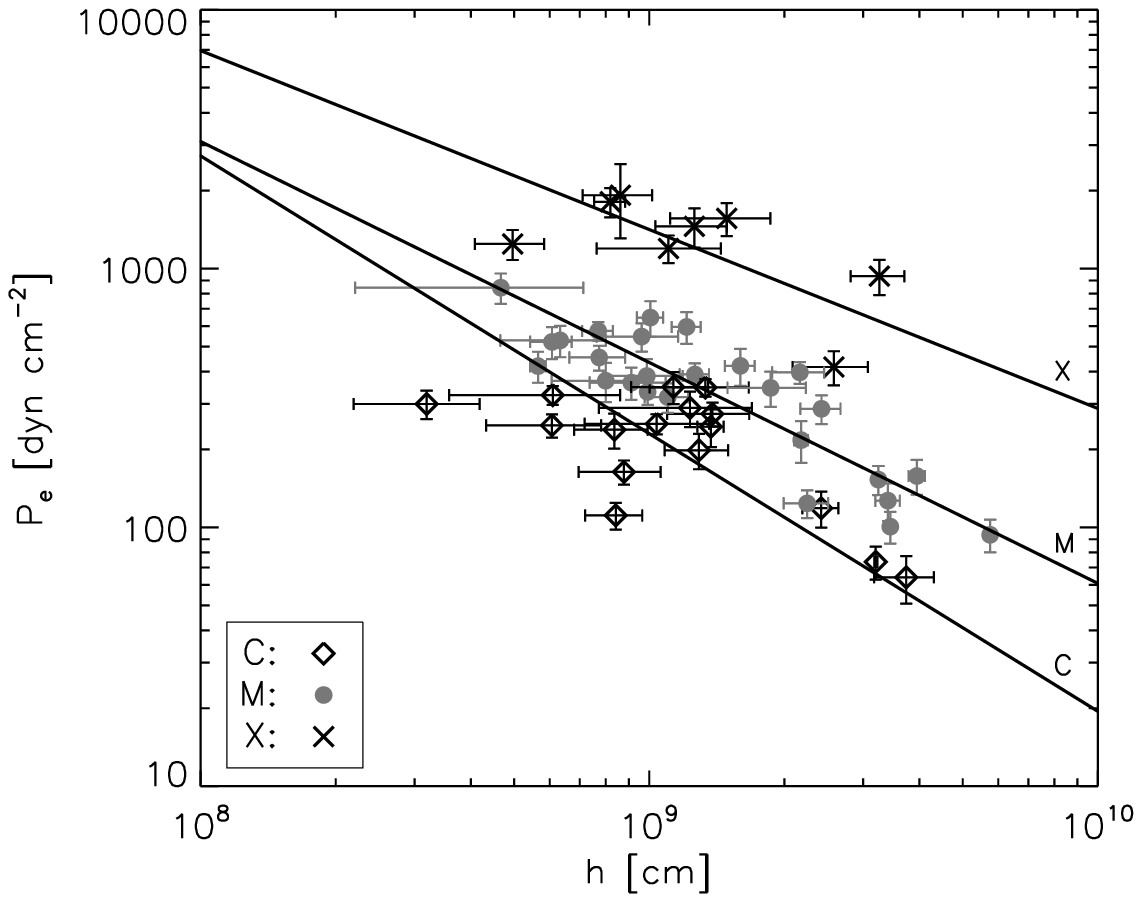}}
\caption{Relation between the plasma pressure in the kernel and its altitude
above the photosphere. The lines show power-law fits to each of three GOES sub-samples.}
\label{p.vs.h}
\end{figure}

Figure \ref{n.vs.h} shows the relation between kernel density and
altitude. Taking the whole sample we can hardly see any relationship,
but if we divide the sample into the GOES classes the relation becomes
more obvious. In a given subclass the kernel density drops with rising
altitude. The observed power-law index of this relationship for C, M and
X-class flares is equal $-1.12\pm0.11$, $-0.82\pm0.04$ and
$-0.65\pm0.12$ respectively. The difference between slopes for M and
X-class kernels is barely significant. The kernels of C-class flares
seem to follow a significantly steeper slope, but the reliability of its
estimation is weak. This sub-sample is ill-spaced with height, as it
includes only 3 kernels with altitude higher than $2\times 10^9$ cm. For
such a sample `bootstrap' methods give more reliable estimation of
slope. Using a bootstrap resampling method 
\cite{bootstrap} we receive the same slope but with error 0.60, which
makes differences between all 3 slopes statistically unimportant.
Similar relations are observed between plasma pressure and kernel
altitude. In Fig. \ref{p.vs.h} we observe very similar power-law slopes
for each GOES-class sub-sample (C: $-1.07\pm0.12$, M: $-0.85\pm0.05$, X:
$-0.69\pm0.14$).

These slopes seem to be defined by the observed geometrical properties
of flare kernels. We do not observe any correlation between temperature
or emission measure with kernel size or altitude. In this case plasma
density calculated from $N_e = \sqrt{EM/V}$ should systematically decrease
with kernel altitude because of the increasing kernel volume. The same relation
should be observed for plasma pressure because the temperature in our
sample is also uncorrelated with geometrical aspects of flare kernels.
As mentioned above, the kernel volume rises with altitude as $V\propto
h^{1.69}$, which should result in density or pressure drop with altitude
as $$N_e = (EM/V)^{1/2}\propto h^{-0.85\pm 0.03}$$ and $$p_e \propto N_e
\propto h^{-0.85}.$$ This slope is consistent with the observed
relation in the sub-sample of M-class flares. For C and X-class flares the
agreement is worse, but this is explained by the X-class sub-sample having small number of events,
and the sub-sample of C-class kernel being ill-spaced as mentioned above.

We interpret these relations as nearly identical concerning the slope,
but differing in the intercept parameter. Stronger flares involve more
emission measure within the same volume, i.e. larger flares need
higher densities and pressures.


\section{Energy release within the kernels}

Loop-top kernels are the brightest parts of the soft
X-ray flares and presumably are the places where the plasma heating is
most effective. The next physical parameter we estimated for the flare
kernels was the heating rate during the kernel maximum. Close to flare
maximum, when the energy transport by plasma flow is negligible we can
build a simplified model of the evolution of thermal energy ($E_{th}$)
content as the sum of heating ($E_H$) and cooling processes (radiative
and conductive losses: $E_R$ and $E_C$):

$$\frac{d}{dt}E_{th} =  E_H - E_C - E_R\,.$$

Moreover, at the flare maximum
the changes of thermal energy content are insignificant, so we can
estimate the energy release rate for this moment as 

$$E_H \approx E_C + E_R\,.$$

Radiative loss can be calculated knowing the temperature and plasma
density estimates, $E_R = N_e^2 \Lambda(T)$, where $\Lambda(T)$ is
the radiative loss function taken e.g. from Rosner, Tucker and Vaiana (1978).

We estimate the kernel heating in the way already shown by Jakimiec, Falewicz and Tomczak
 (2002), which we briefly summarize here. We
assume that the kernel is nearly spherical and it is heated by some
process with the mean rate $E_H$. The total flux of thermal energy
flowing out of the kernel may be estimated as $(4\pi/3)R^3(E_H-E_R)$,
where $R$ is the kernel radius and $E_R$ is the mean radiative
loss from the kernel. This flux has to be transported down by thermal
conduction along the both legs of the loop containing the kernel. If we take the 
diameter of this loop to be the same as that of the kernel, then the conductive flux
 in the legs is $F_C=(2/3)R(E_H-E_R)$. Knowing this
value we can integrate the equation of conduction, $$F_C=\kappa_0T^{2.5}
dt/ds,$$ along the loop of semilength $L$, assuming constant $F_C$. The
result is $$0.286T^{3.5}=(F_C/\kappa_0)L,$$ where $T$ is now the
temperature of the kernel. From these two equations we can estimate the
mean heating within the kernel as 
$$E_H = 3.9\times 10^{-7}\,T^{3.5}/RL+E_R.$$ 
The first term on the right hand of this formula is the conductive loss
from the kernel. Its formulation differs than the radiative
losses from the simple loop.
We want to stress that this estimation does not implicitly involve which
process is responsible for the kernel formation.

\begin{figure}[h!]
\resizebox{\hsize}{!}
{\includegraphics{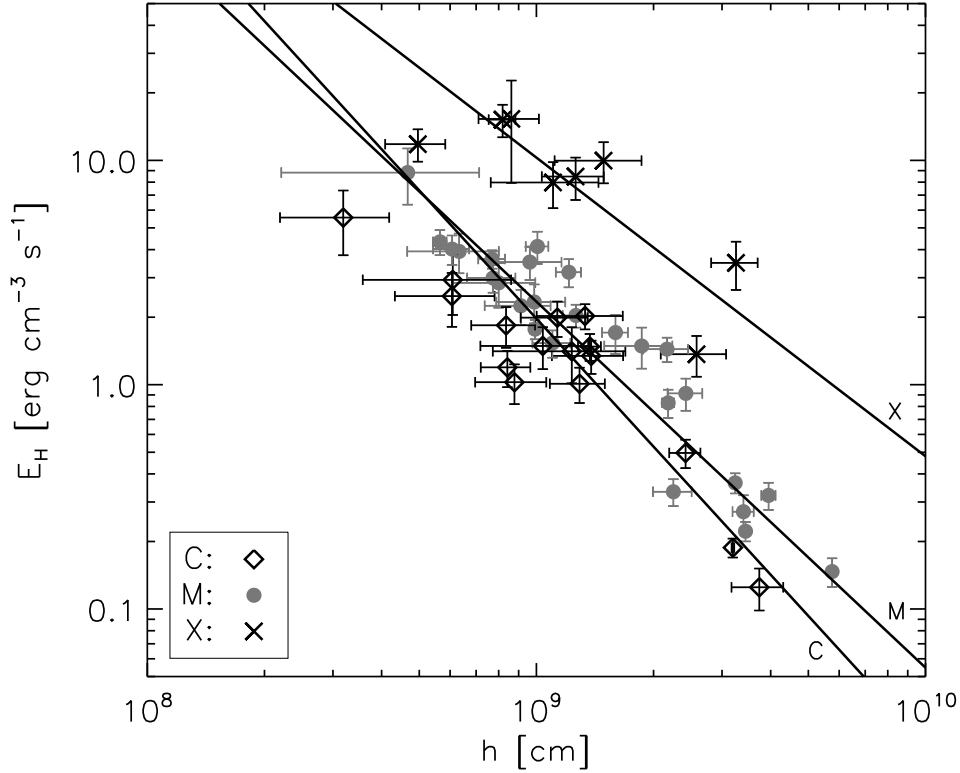}}
\caption{Relation between the energy release rate in the kernel and its
altitude above the photosphere. The lines show power-law fits to each of 
three GOES sub-samples.}
\label{eh.vs.h}
\end{figure}

We applied the above formula to the analyzed sample, taking the
geometrical mean of two semi-axes seen on SXT image as the kernel
radius, and the approximation of the loop semi-length as $L=(\pi/2)\,h$. 
A typical value of heating is a few ergs per second in each cubic
centimeter; however, the observed values within our sample span two
orders of magnitude. Figure \ref{eh.vs.h} shows a clear decrease of the
heating rate with the altitude of a kernel. C and M-class kernels show a
very similar dependence of heating with kernel altitude, $\log E_H =
15.63 -(1.70\pm0.05)\,\log h$. This is similar to the results obtained
by Jakimiec, Falewicz and Tomczak (2002) who analyzed a set of 27 limb
flares of mostly M-class and loop-length longer than $10^9$ cm. They
estimated that the flare heating rate drops with loop-length as $\log
E_H = 13.91 - 1.46\log L$. B\c{a}k-St\c{e}\'slicka and Jakimiec
(2005) enlarged this sample by adding 10 slow long duration
flares, which allowed them to widen the range of analyzed loop-lengths
by an order of magnitude. They obtained the relation $\log E_H = 16.3 -
1.72\log L$. Within errors this is nearly identical to the relation
obtained here for C and M-class flares. Our sample consists also of 8
X-class flares, for which the dependence between heating rate and kernel
altitude is $\log E_H = 13.0 - (1.33\pm0.20)\,\log h$, but the same
slope as in C and M flares cannot be excluded. X-class flares require
about four times higher heating rates, which in our calculations is
caused mainly by distinct increase of emission measure. The radiative
losses in these kernels are correspondingly high to balance the higher
level of heating.

\begin{figure}
\resizebox{\hsize}{!}
{\includegraphics{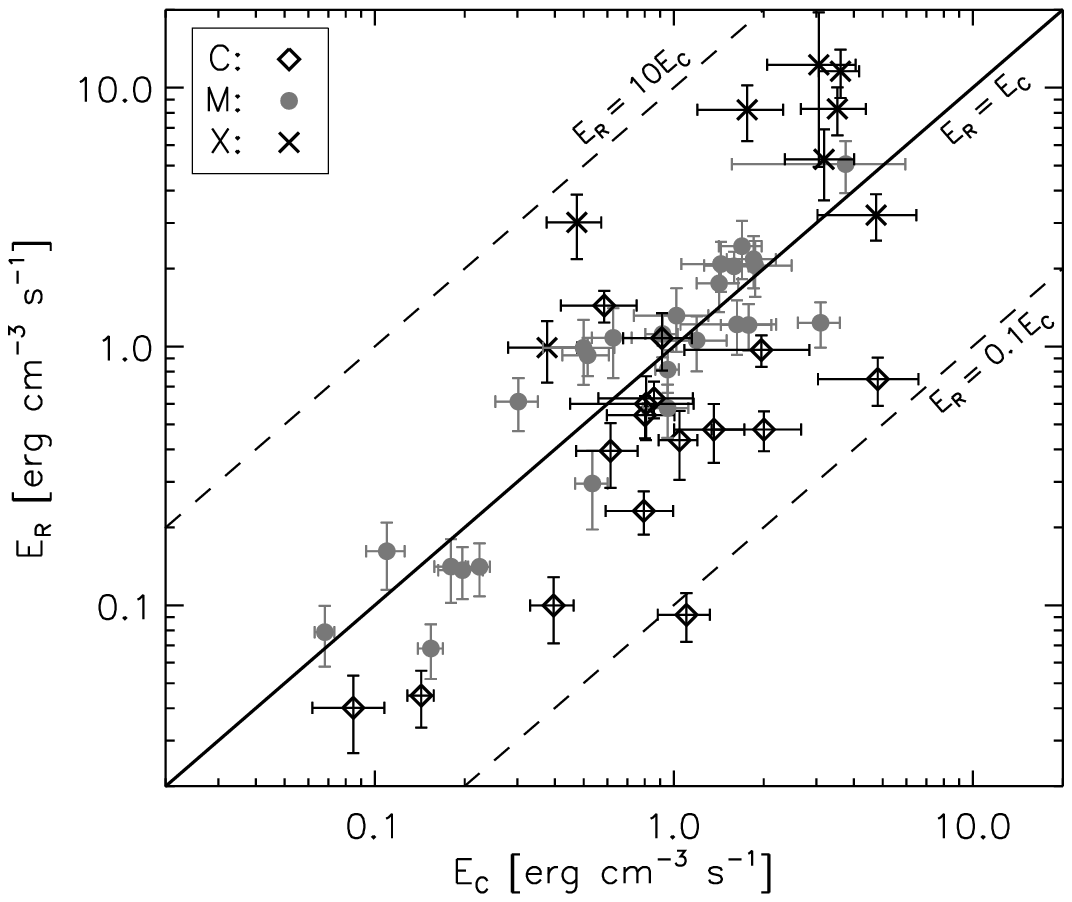}}
\caption{Comparison of the radiative and conductive losses within the analyzed flare kernels.
The solid line shows equality between the two quantities, while the two dashed lines mark
the borders when one loss exceeds the other by a factor 10.}
\label{er.vs.ec}
\end{figure}

These slopes again seem to result mainly from the geometrical
properties of the kernels. During the flare maximum radiative and
conductive losses are usually comparable, $E_C \sim E_R$ (see
Fig. \ref{er.vs.ec}). With the lack of any dependence of 
temperature on kernel altitude, the conductive losses scale as
$E_C \sim (RL)^{-1}$. Taking our result, $R\sim L^{0.56}$, we obtain
$E_H \sim E_C \sim L^{-1.56}$. This slope differs from the relations
in Fig. \ref{eh.vs.h} by no more than $3\sigma$.


Figure \ref{er.vs.ec} also shows that the ratio between the radiative
and conductive losses changes when going from weak to strong flares. The
stronger the flare, the more important the influence of radiative losses.
Conductive losses depend on the geometrical aspects and the plasma
temperature. The geometrical properties of flare kernels are more or
less the same for each GOES sub-sample. Taking the relation between
temperature and emission measure observed in our sample we can show that
 $E_C\propto T^{3.5}/RL\propto
EM^{0.315}/h^{1.56}$. Radiative losses are proportional to emission measure,
$E_R=EM\times\Lambda(T)/V\propto EM^{0.94}/h^{1.69}$, where we
assumed that $\Lambda(T)\propto T^{-2/3}$. Both losses have similar
dependence on the kernel altitude, but different on the
emission measure, which makes radiative cooling more important in strong
flares, $E_R/E_C\propto EM^{0.63}/h^{0.13}$. This result is, however,
sensitive to the assumed dependence between temperature and emission
measure. The SXT instrument has limited sensitivity to the hot
plasma which is the cause of the quite flat relation between these
two parameters in our sample. It is enough to assume relation $T\propto
EM^{0.24}$, which is power-law equivalent of the relation presented by Feldman
et~al. (1996), to get the ratio $E_R/E_C$ independent
of the kernel emission measure. 

The limited temperature sensitivity of the SXT affects both radiative and
conductive cooling. Underestimation of kernel temperature results in
overestimation of radiative and underestimation of conductive losses.
The second effect is stronger, because it depends on temperature to a much
higher power. Thus the heating rate calculated in this paper seems also
to be underestimated and should be treated as a lower limit. This bias
should be more important for stronger flares. A typical value of
kernel temperatures for X-class flares in our sample is $T\approx 12$ MK,
which seems to be a factor 2 less than temperatures reported by other
similar instruments. This would result in underestimating of
conductive losses by about one order of magnitude and overestimating the
radiative losses by a factor 1.6. Good estimation of cooling and heating
rates for strong flares requires an imaging X-ray telescope with better
temperature sensitivity than SXT.

\section{Discussion and conclusions}

The observed relation between kernel size and its altitude is not an
obvious result. If we assume that kernel is part of the loop that
appears to contain it, than the loop cross-section area must show the
same relation. This means that the higher flare loops are not linearly
scaled version of the smaller loops, i.e. the loop diameter,
$\Phi\approx(4A/\pi)^{1/2}$ is not proportional to the loop length, but
rather to its square root, $\Phi\sim h^{1.13/2}\sim L^{0.56}$. The
higher loops must be systematically more narrow than the smaller ones.

The relation between plasma pressure and the kernel altitude allows us
to put some constrains on the magnetic field structure in active
regions. The kernel plasma of a given pressure must be trapped within a
field with magnetic pressure greater than plasma pressure, $B^2/8\pi >
p_e$. This allows us to determine the minimum magnetic strength necessary
to contain the kernel plasma as $\log B_{min} = 5.86 - 0.43 \log h$ for
M-class kernels. This constraint slightly depends on the GOES class. If
we assume that for all flares $\log p_e$ decreases with altitude with the
same slope as for M-class kernels, the intercept
parameter of $\log B_{min}$ in C-class flares is 5.71 while in X-class flares it is 6.11.
This estimation does not take into account the influence of possible
field helicity, which allows a weaker field effectively to trap a
kernel plasma. The field helicity is unfortunately difficult to determine with
present X-ray observations of flares. 

\begin{figure}
\resizebox{\hsize}{!}
{\includegraphics{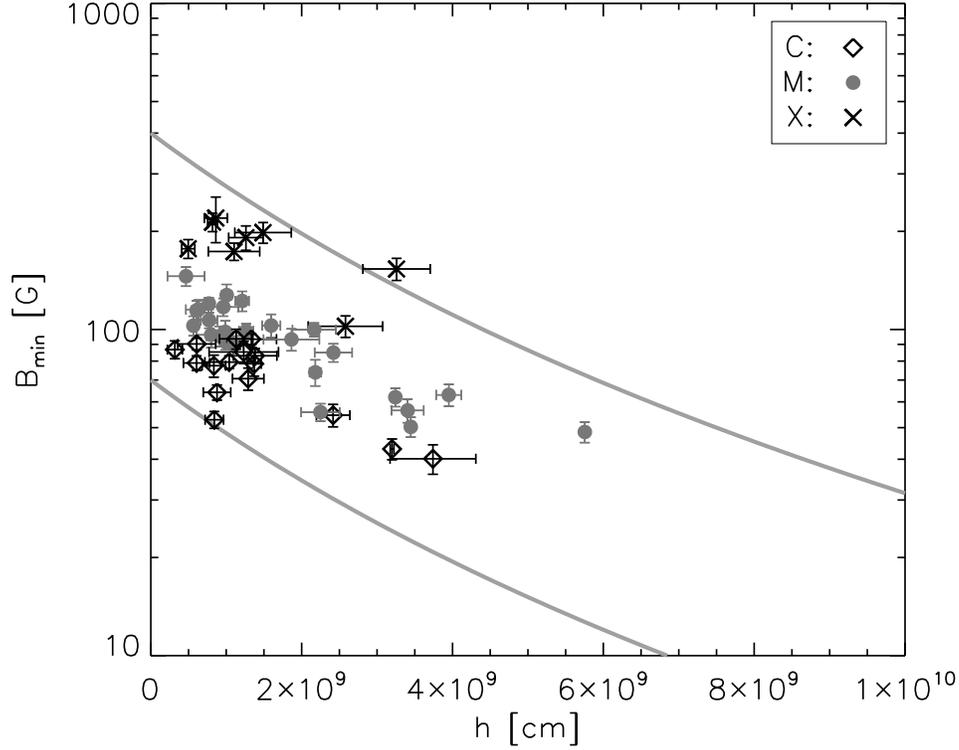}}
\caption{Relation between the magnetic field necessary to trap the
kernel plasma and the kernel altitude. The two lines presents the dipole field, 
according to Aschwanden et~al. (1999) approximation with 
$h_D=7.5\times 10^9$ cm and $B_0$ equal 70 and 400 G.}
\label{b.vs.h}
\end{figure}

The observed relation $B_{min}(h)$ is also in quite good agreement with
estimates of magnetic field in active regions shown by Aschwanden
et~al. (1999). They approximated the field with a magnetic dipole,
$B(h)\approx B_0 (1+h/h_D)^{-3}$, where $h_D$ is the dipole depth
estimated by them as $7.5\times 10^9$ cm, and $B_0$ is the mean magnetic
field at the photospheric level. Fig. \ref{b.vs.h} shows that our sample
is consistent with the dipole field with $B_0$ in the range of 70 to 400
G.

The relation obtained between kernel heating and its altitude may also
shed light on the magnetic field structure in active regions. If we
hypothetically assume, as in Jakimiec, Falewicz and Tomczak (2002), that
the energy release rate is somehow proportional to the density of
magnetic energy $E_H \sim B^2/8\pi$, then we should observe decrease of
the magnetic field with height as $B \sim h^{-0.85}$. Jakimiec, Falewicz
and Tomczak (2002) suggested in the same way that magnetic field
responsible for flare heating decreases with height as $B\sim
h^{-0.73}$. This relation has a different slope from the decrease of
magnetic field necessary to trap the kernel plasma, but both relations
may describe different parts of the field supporting the kernel. The
field necessary to contain the kernel plasma may be more related to the
kernel surroundings, while the field responsible for heating is
obviously related to the location of energy release. These two places
are not by definition the same location as it is discussed e.g. in
Karlick\'y and Bart\'a (2006).

The reported here relation between the kernel size and its altitude
should be considered regarding the flare scaling laws. It is
generally accepted that during the flare maximum radiative and
conductive losses are of the same order, $E_R\sim E_C$. Conductive losses
for a simple flaring loop without a kernel, $E_C \propto T^{3.5}/L^2$, changes into
$E_C\propto T^{3.5}/RL$ when we describe a loop-top kernel. In this
case the scaling takes the form $N_e^2\Lambda(T)\sim
T^{3.5}/RL$. If we substitute $\Lambda(T)\sim T^{-\beta}$, then we can
rewrite this scaling as $$N_e^2RL\sim T^{3.5+\beta}.$$ The 
relation between kernel size and its altitude is $R\sim h^{0.56}$,
which makes the product $RL\sim h^{1.56}$ close to kernel volume
$R^3\sim h^{1.68}$. Thus, the left side of the above equation roughly scales 
as kernel emission measure. The scaling $E_R\sim E_C$ turns into
a relation between flare temperature and emission measure,
$EM\sim T^{3.5+\beta}$. Taking $\beta=2/3$ we achieve $T\sim EM^{0.24}$
which is very close to the relation shown by Feldman et~al.
(1996).

The existence of flare kernels should be taken into account also when
analyzing stellar flares. It is hard to imagine that in stellar
flares loop-top kernels are not present when in solar ones they are
common. We should be careful when transferring the scaling laws of solar
kernels onto flares on other stars. Recognizing how the overall magnetic
structure affects formation and evolution of a flare kernel is
necessary. The Sun represents stars with very low magnetic
activity. We cannot exclude the possibility that the kernel scaling laws may have
different formulation on highly active stars, which seems to be fully
covered by active regions. However, we remark that the similarity of a $T-EM$
relation for stellar flares with a solar one (see e.g. Fig. 1 in
Shibata and Yokoyama, 2002) suggests that kernel scaling laws on
other stars may not differ substantially.

Flare loop-top kernels are obvious characteristics of SXR solar flares.
The kernel emission is usually constant part of the total flare
emission, what allows to describe overall evolution of a flare by the
evolution of its kernel(s). A correct model of either solar or stellar
flares cannot be built without understanding their properties and
nature.


This study shows that the set of basic observables describing the flare
loop-top kernels, which we obtain from SXT images, separates into two
independent pairs. One pair describes the geometrical properties of a
kernel (size and altitude), the other pair is plasma temperature and
emission measure. As we report here, kernel size and altitude are not
independent parameters. Although in individual flares one can see
different types of relation between them, when comparing many flares, a
clear dependence emerges: the higher the kernel, the bigger the size.
The second pair, temperature and emission measure, is also internally
related, but these parameters are not correlated with the kernel
altitude and size. To describe many flares, it is sufficient to take
only two observables, viz. emission measure and kernel height. Other
physical parameters of emitting plasma like density, pressure or heating
rate are related to these two quantities. 

To determine more precisely the kernel scaling laws, we need a larger
sample of limb flares. We expect that the forthcoming imaging instrument in
soft X-rays, $XRT$ onboard {\em Solar-B}, will open new possibilities in
analyzing the physics of flare loop-top kernels. The expected improved angular
resolution will allow better determination of kernel geometrical
properties. Improved thermal sensitivity of $XRT$ should allow us to
achieve more precise determination of physical properties of the emitting
plasma. We may expect better determination of conductive losses and
heating rates, which strongly depend on the estimated kernel
temperature. The next few years of soft X-ray observation of the Sun
will allow us to substantially enlarge the sample of suitable,
near-limb flares and take a closer look on the kernel properties.
Difficult to model, flare kernels still await our attention.

\section*{Acknowledgments}
The {\em Yohkoh} satellite was a project of the Institute of Space and
Astronautical Science of Japan.
This work has been supported by the Polish Ministry of Science and High
Education grant N203 001 32/0036.


\begin{thebibliography}{}

\bibitem[\protect\citeauthoryear{Acton et al.}{1992}]{acton}
  Acton L.\,W., Feldman U., Bruner M.\,E., et~al.: 1992, {\it PASJ} {\bf 44}, L71.
\bibitem[\protect\citeauthoryear{Aschwanden et al.}{1999}]{aschi}
  Aschwanden M.\,J., Newmark J.\,S., Delaboudini\`ere J.-P., et~al.: 1999, {\it Astrophys. J.} {\bf 515}, 842.
\bibitem[\protect\citeauthoryear{B\c{a}k-St\c{e}\'slicka and Jakimiec}{2005}]{ula}
  B\c{a}k-St\c{e}\'slicka U., Jakimiec. J.: 2005, {\it Solar Phys.} {\bf 231}, 95.
\bibitem[\protect\citeauthoryear{Bentley et al.}{1994}]{bentley}
 Bentley R.\,D., Doschek G.\,A., Simnett G.\,M. et~al.: 1994, {\it Astrophys. J.} {\bf 421}, L55.
\bibitem[\protect\citeauthoryear{Doschek, Strong and Tsuneta}{1995}]{doschek}
  Doschek G.\,A., Strong K.\,T., Tsuneta S.: 1995, {\it Astrophys. J.} {\bf 440}, 370.
\bibitem[\protect\citeauthoryear{Feldman et al.}{1995}]{feldman95}
  Feldman U., Seely J.\,F., Doschek G.\,A., et~al.: 1995, {\it Astrophys. J.} {\bf 446}, 860.
\bibitem[\protect\citeauthoryear{Feldman et al.}{1996}]{feldman}
  Feldman U., Doschek G.\,A., Behring W.\,E., Phillips K.\,J.\,H.: 1996, {\it Astrophys. J.} {\bf 460}, 1034.
\bibitem[\protect\citeauthoryear{Hara et al.}{1992}]{filter.method} 
  Hara H., Tsuneta S., Lemen J.\,R., Acton L. W., McTiernan J.\,M.: 1992, {\it PASJ} {\bf 44}, L135.
\bibitem[\protect\citeauthoryear{Jakimiec}{2002}]{jakimiec02j}
  Jakimiec J.: 2002, {\it Adv. Space Res.} {\bf 30}, 577.
\bibitem[\protect\citeauthoryear{Jakimiec et al.}{1998}]{jakim-ker}
  Jakimiec J., Tomczak M., Falewicz R., Phillips K.\,J\,.H., Fludra A.: 1998, {\it Astron. Astrophys.} {\bf 334}, 1112.
\bibitem[\protect\citeauthoryear{Jakimiec et al.}{2002}]{jakimiec02}
  Jakimiec J., Falewicz R., Tomczak M.: 2002, {\it Adv. Space Res.} {\bf 30}, 659.
\bibitem[\protect\citeauthoryear{Karlick\'y and Bart\'a}{2006}]{karlicky}
  Karlick\'y M, Bart\'a M.: 2006, {\it Astrophys. J.} {\bf 647}, 1472.
\bibitem[\protect\citeauthoryear{K\c{e}pa et al.}{2005}]{ania} K\c{e}pa A., Sylwester J., Sylwester B.,
Siarkowski, M., Kuznetsov, V.: 2005, in Proceedings of the 11th European
Solar Physics Meeting {\bf ESA SP-600}, ed. D. Danesy, S. Poedts, A. De Groof \&  J. Andries, 87.1.
\bibitem[\protect\citeauthoryear{Masuda}{1994}]{masuda}
  Masuda S.: 1994, Ph.\,D. Thesis, University of Tokyo.
\bibitem[\protect\citeauthoryear{Press, Teukolsky and Vatterling}{1992}]{bootstrap}
  Press W.\,H., Teukolsky S.\,A., Vatterling W.\,T.: 1992, in {\it Numerical Recepies in Fortran}, 
  2nd ed., Cambridge and New York, Cambridge University Press.
\bibitem[\protect\citeauthoryear{Rosner, Tucker and Vaiana}{1978}]{rtv}
  Rosner R., Tucker W.\,H., Vaiana G.\,S.: 1978, {\it Astrophys. J.} {\bf 220}, 643.
\bibitem[\protect\citeauthoryear{Shibata and Yokoyama}{2002}]{shibata} 
  Shibata K., Yokoyama T.: 2002, {\it Astrophys. J.} {\bf 577}, 422.
\bibitem[\protect\citeauthoryear{Tomczak}{1997}]{tomczak} 
  Tomczak M.: 1997, {\it Astron. Astrophys.} {\bf 317}, 223.
\bibitem[\protect\citeauthoryear{Tsuneta et al.}{1991}]{sxt} 
  Tsuneta S., Acton L., Bruner M. et~al.: 1991, {\it Solar Phys.} {\bf 136}, 37.
\bibitem[\protect\citeauthoryear{Vorpahl, Tandberg-Hanssen and Smith}{1977}]{vorpahl}
  Vorpahl J.\,A., Tandberg-Hanssen E., Smith J.\,B.: 1977, {\it Astrophys. J.} {\bf 212}, 550.
\bibitem[\protect\citeauthoryear{Yokoyama and Shibata}{1998}]{yokoyama}
  Yokoyama T., Shibata K.: 1998, {\it Astrophys. J.} {\bf 494}, L113.


\end{thebibliography}

\end{article}
\end{document}